\documentstyle[twoside,fleqn,epsf,espcrc2]{article}

\newcommand{\AmS}{{\protect\the\textfont2
  A\kern-.1667em\lower.5ex\hbox{M}\kern-.125emS}}

\input epsf.sty
\title{
\vspace{-1.5cm}
\hfill \normalsize{OCHA-PP-142} \\
\vspace*{0.3cm}
Landau Gauge Fixing supported by Genetic Algorithm}
\author{Azusa Yamaguchi\address{Particle Physics Lab.Department of Physics,
Ochanomizu University,\\
2-1-1 Otsuka, Tokyo 115 Japan (e-mail azusa@sokrates.ocha.ac.jp)} and
Hideo Nakajima\address{Department of Information Science, 
Utsunomiya University,\\
2753 Ishii, Utsunomiya 321-8585 Japan (e-mail nakajima@is.utsunomiya-u.ac.jp)} 
}
\begin{document}
\begin{abstract}
A class of algorithms for the Landau gauge fixing is proposed, 
which makes the steepest ascent (SA) method be more efficient by
concepts of genetic algorithm.
Main concern is how to incorporate random gauge transformation (RGT)
to gain higher achievement of the minimal Landau gauge fixing,
and to keep lower time consumption. One of these algorithms uses 
the block RGT, and another uses
RGT controlled by local fitness density, and the last
uses RGT determined by Ising Monte Carlo process.
We tested these algorithms on SU(2) lattice gauge theory
in 4 dimension with small
 $\beta$s, $2.0, 1.75$ and $1.5$,
and report improvements in hit rate and/or
in time consumption, compared to other methods.
\end{abstract}

\maketitle

\begin{table*}[tb]
\setlength{\tabcolsep}{1.5pc}
\caption{ 
 Search parameters and strategies on SU(2)with $\beta=2.0$ 
performed on DEC $\alpha$ 2100 4/275.
\label{tab:SP}}
\begin{tabular*}{\textwidth}{llccc}\hline
Method & &Hitrate & executing time [sec]\\
\hline 
BRGT& IS on $N_{div}=2$ & $46/50$  &  $10.9$ \\
    & IS off $N_{div}=2$ & $42/50$ &  $11.6$ \\
\hline 
IRGT & IS off & $32/50$  & $11.2$ \\
\hline
LFRGT& $R1=0.5$,$R2=0.85$ & $47/50$  & $11.6$ \\
     & $R1=0.35$,$R2=0.85$ & $46/50$  & $14.1$ \\
     & $R1=0.45$,$R2=0.9$ & $45/50$  & $10.3$ \\
     & $R1=0.5$,$R2=0.85$ IS on& $44/50$  & $12.7$ \\
\hline 
HdeF & & $0/50$ &$7.6$ \\ 
\hline
TRGT & & $47/50$ & $14.7$ \\
\hline
\end{tabular*}
\end{table*}
\section{INTRODUCTION}

Gauge fixing degeneracies (existence of Gribov's copies)
are generic phenomena\cite{Gv,Sg,MN}
in non-abelian continuum gauge theories
, and thus there exist fundamental problems, e.g.,
what is the correct gauge fixed measure in the path integral formalism.

Although the foudation of lattice gauge theories does not necessitate
gauge fixings, they are, however, often requeired for 
field theoretic transcriptions of their nonperturbative dynammics.

The principle of the Landau gauge fixing algorithm
is given as an optimization problem of some functions along
the gauge orbit of
link variables\cite{MN}.
There are many extrema of the optimization function in general,
and all these extrema points on the gauge orbit correspond to
the Landau gauge, Gribov copies. The absolute maximum
corresponds to the
minimal Landau gauge\cite{Zw}.
In order to fix the Landau gauge
uniquely, the minimal Landau gauge is the most favourable goal.

Local search algorithms were developed by a chain of gauge
transformations\cite{MO,dFG}.
.
Since there are
many Gribov copies,
the gauge orbit paths are easily captured by these extrema.
We call these method as the steepest ascent
method (SA). In such a situation that
SA fails to attain an absolute extremum,
one can try a simple random search\cite{Cu} on the gauge orbit
with succeeding SA.

There is only unique
method of Hetrick-de Forcrand\cite{HdF}
 (HdeF) aiming
at the minimal Landau gauge fixing,
which is systematic in the sense that random trial is not involved.
However it works successfully only for large $\beta$ samples, rather
smooth configurations.
Thus finding efficient algorithm
for the minimal Landau gauge fixing is still an open problem.

We report results of an attempt in some GA type methods 
in comparison with other methods as the simple RGT method.
We work on  $SU(2)$ gauge theory of $8^4$ lattice, and
define gauge fields as $
A_{x,\mu}=\displaystyle{1\over 2i}(U_{x,\mu}-U^{\dag}_{x,\mu})$,
then the optimization function, fitness, is given as
\begin{equation}
F_U(G)=\sum_{x,\mu}\displaystyle{1\over 2}Tr(U^G_{x,\mu}),
\end{equation}
where $U^G_{x,\mu}=G^{\dag}_x U_{x,\mu}G_{x+\mu}$.
The fitness $F_U(G)$ can be viewed as
negative of energy of $SU(2)$ spin system $G$ sitting on
sites, and 
thus the problem is equivalent to finding the lowest energy
state under the randomized interaction $U$.

Straightforward GA strategy was applied to
the minimal Landau gauge fixing\cite{MK}. Their preliminary results
of the Landau gauge fixing
show that the straightforward application of GA is not so good
as to become a practical method.

We tested three types of GA methods, which are different from each other
in how RGT are incorporated in the algorithms.
We compare them with the simple RGT method in performance. 

\section{Algorithms}

Our aim is to develop algorithms efficient for small $\beta$s.
Basic building block for the algorithms
is RGT. The simple RGT (TRGT) method
which applies RGT on the whole lattice takes time,
while it is difficult for SA paths to escape from a Gribov copy 
if the RGT is restricted in blocks of too small area on the lattice.

Thus main features of our algorithms consist in how to determine blocks
to which RGT is applied.
Given a Gribov copy link configuration $U$, the following three types
of definition of blocking for RGT are devised.

1.
The whole lattice is partitioned into 
${N_{div}}^{d}$ chequered blocks, where the number of dimension, $d=4$.
Then RGT is applied on white blocks and
a constant RGT on each black one, and vice versa. We call this method
as blocked RGT (BRGT) method.

2.
We set two parameters $R1$ and $R2$, and
sites $i$ where RGT is applied are chosen according to local
fitness density, $f(i)$, by $f(i)<R1$ or $R2<f(i)$. We call this method
as local fitness density RGT (LFRGT).

3.
Ising spin interaction on randomly chosen coarse lattice is defined
from gauge spin interaction given by $U$ such that at least one antiferro
interaction should be involved. Through Monte Carlo simulation
of this Ising system, one obtains up-spin blocks $B_{+}$ and down-spin blocks
$B_{-}$.
Blocks for RGT is given by one of these blocks. $\beta_{Ising}$ is
so chosen that size of both blocks $B_{\pm}$ becomes comparable.
We call this method as Ising RGT (IRGT).

We use the local exact algoritm\cite{MO,dFG} for SA method.
Given a Gribov copy $U$, the SA method following
RGT in use of one of three blockings, brings $U$ to
the extrema of fitness by steps of gauge transformations.
This new copy in the Landau gauge is put as an initial
copy for the next iteration.
This sub-procedure is repeated $M_{itr}$ times. 
The original Gribov copy and 
the maximum fitness copy among $M_{itr}$ new gauge copies
are compared in fitness value.
If the fitness of the new one is higher, the sub-procedure is
to be started again with this new one as an initial copy.
Otherwise the process stops, and the initial copy is considered as
the expected configuration with the maximum fitness value.

In addtion to the above basic algorithm, two types of modification
are devised as follows:

1.As the initial copy in the sub-process,
the better fitness copy between the current copy and the preceding, is always
chosen.
We call this as the inter-selection-on (IS) scheme.

2.As the initial copy in the sub-process, the product of crossing
is adopted, where chequered block crossing is done between
the best fitness copy and the
second best one among the obtained copies so far.
We call this as the crossing-on (C) scheme.

We tested these algorithms on $SU(2)$ $8^4$ lattice with $\beta=2.0,1.75,
1.5$
and tuned some parameters, as $N_{div}$, $\beta_{Ising}$ and $M_{itr}$,
with or without IS and/or C scheme.

\section{Results and Performance}
Our three GA type methods were executed on the same set of randomly
produced 50 copies from a suitably chosen copy from
samples, $\beta=2.0,1.75,1.5$.
The TRGT method
and the method of HdeF were also tested on the same set.
Performance of these methods, hit rate of the minimal Landau gauge and
average time consumption for the gauge fixing, are compared.

We fix $M_{itr}=5$, and in tests of parameters search for $\beta=2.0$,
 we found that
BRGT with $N_{div}=2$ shows a sufficient global search power,
while BRGT with $N_{div}=4$ shows a lower hit rate.
The IS scheme could be viewed as a kind of elitism, and
it is known that elitism is a suitable strategy when the global
search power is available.
BRGT with IS scheme, $N_{div}=2$, shows a powerful and efficient search, while
IRGT with IS, however, does not. 
For LFRGT, the cut parameters, $R1$ and $R2$, affect its 
search power. Without $R2$ or with too large $R2$,
even high $R1$ does not work well, nor low $R2$ without $R1$.
Since LFRGT with both paramters suitabley chosen, 
achieves the efficient search, IS scheme helps.
From Table~\ref{tab:SP}, our algorithms, BRGT with IS, $N_{div}=2$, and
LFRGT with IS, $R1=0.5$ and $R2=0.85$, have high hit rates and
exhibit good performance comparable with TRGT. The HdeF method with
small $\beta$ is known that it does not tend to the maximun fitness~\cite{HdF}.

\begin{figure}[tb]
\begin{center}
\vspace{-1.0cm}
\epsfxsize=7.5cm\epsffile{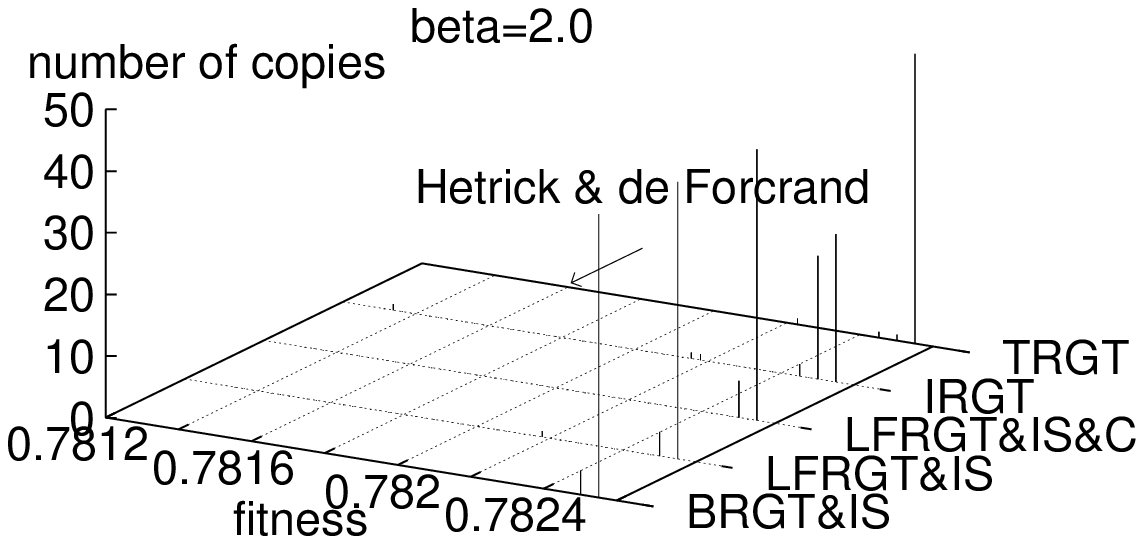}
\vspace{-1.7cm}
\epsfxsize=7.5cm\epsffile{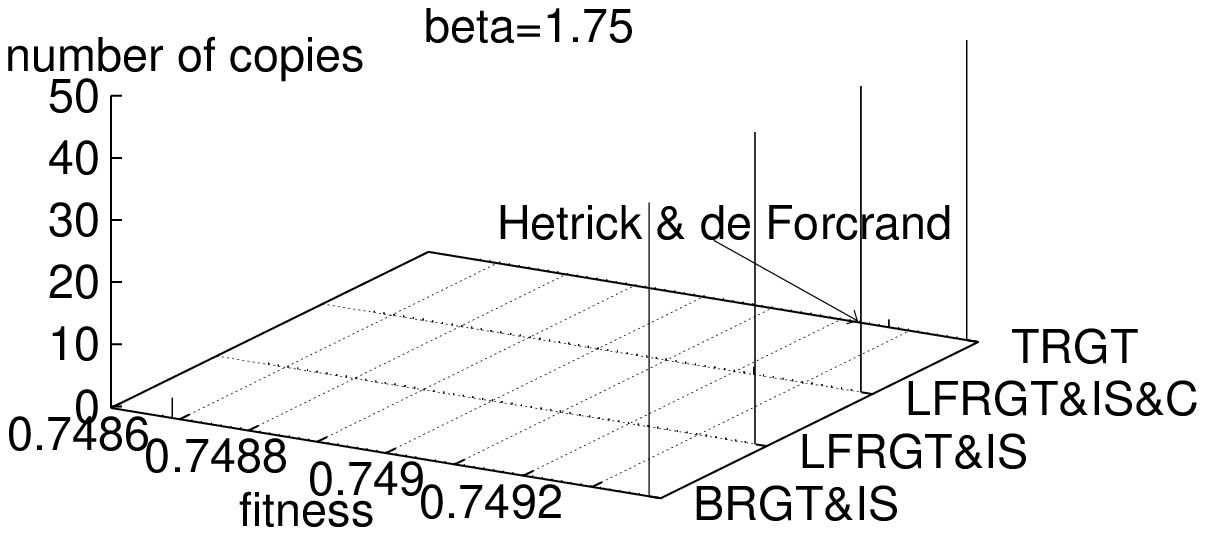}
\vspace{-1.7cm}
\epsfxsize=7.5cm\epsffile{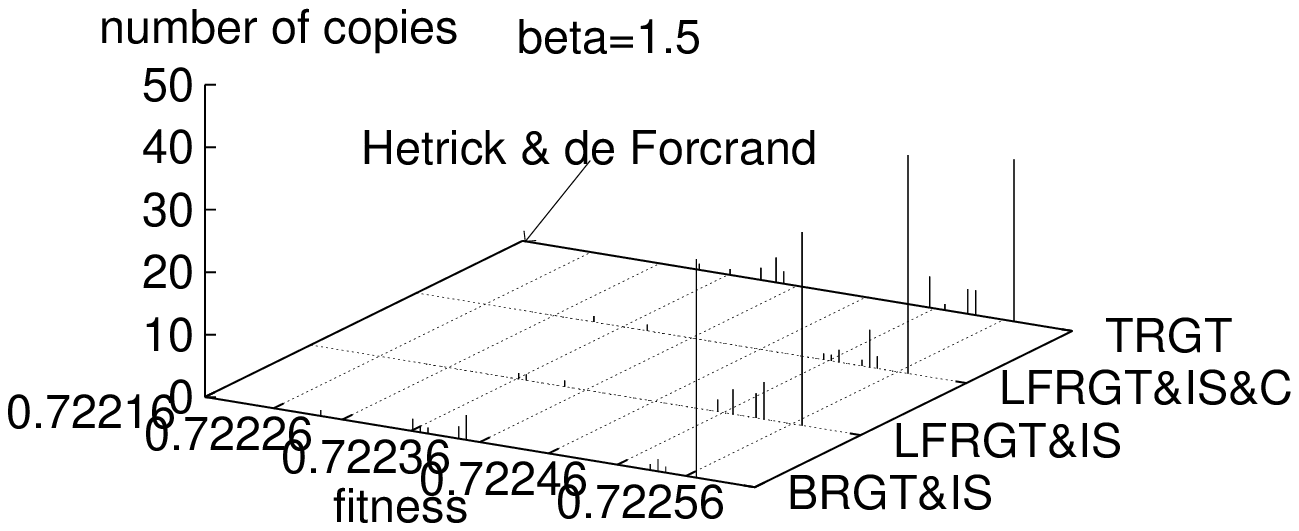}
\vspace{-1.7cm}
\end{center}
\caption{ \label {fig:HR}
Gribov copies obtained by the various algorithms and the
{\it unique copy} $\downarrow$ obtained by HdeF algorithm.}
\end{figure}

The hit rate performance for various methods is given in Figure~\ref{fig:HR}, 
and the average time consumption is shown in Table~\ref{tab:SP}.

This work is supported 
by Japan Society for the Promotion of Science, Grant-in-aid for Scientific 
Research(C) (No.11640251).

\end{document}